\documentclass[11pt]{article}
\begin{document}
\centerline{\Large Bohr, objectivity, and ``our experience''}
\smallskip
\centerline{\large\textit{\`A propos} Mermin's note on the quantum measurement problem}
\bigskip
\centerline{\sc Ulrich J. Mohrhoff}
\medskip
\begin{abstract}
In a recent note David Mermin attributed the idea that wave function collapse is a physical process to a misunderstanding of probability and the role it plays in quantum mechanics. There are, however, further misconceptions at play, some of which are shared by Mermin himself and more generally by QBists. The main objective of the present comment is to explain why I disagree with his reading of a well-known passage by Niels Bohr, in particular the ambiguity of the first person plural he perceives in Bohr's reference to ``our description of nature'' and ``our experience.''
\end{abstract}
\section{Introduction}
In a recent note\cite{Mermin2022} David Mermin attributed the idea that wave function collapse is a physical process to a misunderstanding of probability and the role it plays in quantum mechanics.%
\footnote{Implicit in Mermin's critique of the collapse of a quantum state is a critique of the notion that a quantum state's dependence on time is the time dependence of an evolving state, rather than a dependence on the time of the measurement to the possible outcomes of which a quantum state serves to assign probabilities.}
There are, however, further misconceptions at play, some of which are shared by Mermin himself and more generally by proponents of QBism, to which Mermin refers here as ``the interpretation of quantum mechanics by Carlton Caves, Christopher Fuchs, and R\"{u}diger Schack.'' 

My main objective in this comment is to explain why I disagree with Mermin's reading of the  following frequently quoted passage by Niels Bohr\cite{Bohr1934}: ``[i]n our description of nature the purpose is not to disclose the real essence of the phenomena but only to track down, so far as it is possible, relations between the manifold aspects of our experience.'' Mermin holds that the possessive pronoun in ``our description of nature'' and ``our experience'' is to be read ``not as all of us collectively but as each of us individually,'' and he believes that the ``unacknowledged ambiguity of the first person plural lies behind much of the misunderstanding that still afflicts the interpretation of quantum mechanics.''

Elsewhere\cite{MerminQBnotCop} Mermin has pointed out that ``[s]cience is a collaborative human effort to find, through our individual actions on the world and our verbal communications with each other, a model for what is common to all of our privately constructed external worlds.'' In what follows I will show that Bohr does \textit{not} refer to the description or the experience of our respective privately constructed external worlds but in fact to the description or the experience of that part of our privately constructed external worlds which is common to all of them and therefore to all of us.

\section{Self-evident truths}
The unsuspecting public may be excused for believing that the predictions of quantum mechanics are based on a model of reality, and that the accuracy of the theory's predictions testifies to the model's correctness. But such is not the case. Quantum mechanics, and more generally physics, offers us calculational tools. In pre-quantum physics it \textit{seemed} possible to consistently transmogrify certain calculational tools into (descriptions or representations of) actual physical entities or processes. One could pretend, for example, that the electromagnetic field not only serves to calculate the effects that electrically charged objects have on electrically charged objects but also describes the process (or mediates the action) by which charges act on charges.\cite{Mermin2009} 

With Mermin, I hold these truths to be self-evident, that quantum physics equips us with mathematical tools for assigning probabilities to the possible outcomes of measurements that may be made, on the basis of information provided by the outcomes of measurements that have been made, and that attempts at transmogrifying such a tool into a physical entity or process are reduced to absurdity by the numerous calamities to which they lead, such as ``the disaster of objectification''\cite{vF90} or ``von Neumann’s catastrophe of infinite regression'',\cite{dW70} not to mention the existing proofs of insolubility theorems for the objectification problem.\cite{Mittelstaedt98,BLM96}. 

A not insignificant number of physicists nevertheless believe that the outcome of an earlier measurement exerts some kind of influence on a reified wave function, which then evolves as dictated by a Schr\"odinger equation or some relativistic analogue, until it exerts some kind of influence on a later measurement, which determines the probabilities of its possible outcomes.

The perceived need for a medium or mediating process can be supported by a tale of considerable appeal. It goes like this. Since the past is no longer real, it can influence the present only through the mediation of something that persists through time. Causal influences (including statistical ones) reach from the no longer existing past into the not yet existing future by being ``carried through time'' by something that ``stays in the present.'' There is, accordingly, an evolving state, and this includes everything in the past that is causally or statistically relevant to the future. That is how, in classical physics, we come to conceive of physical fields that serve as mediums by which past causes produce future effects. It is also how, in quantum physics, people come to believe that the wave function plays a similar mediating role.

Today we know from the violation of Bell's inequality and other “no-go theorems” that the synchronic correlations predicted by quantum mechanics cannot be explained by any kind of mediating process. Why then should the diachronic correlations predicted by quantum mechanics be any different? (Synchronic correlations obtain between the outcomes of measurements performed on different systems at one and the same time; diachronic correlations obtain between the outcomes of measurements performed on one and the same physical system at different times.) Although the founders of the quantum theory could not have been aware of such theorems, most had the good sense not to think of the time-dependent wave function as (representing) a medium or a mediating process. What made it rather easy for them to forbear such a notion was that quantum mechanics \textit{presupposes} the occurrence of the events which it serves to correlate; it accounts neither for their existence nor for the classical world in which they occur.

With Mermin, I also hold this truth to be self-evident (though it took me some time to get there), that probabilities are intrinsically subjective. There is---as there always is---a contrary view according to which probabilities are subjective only if, and only because, they are assigned on the basis of incomplete information. On this view, quantum mechanics is taken to imply that probabilities are not necessarily associated with ignorance or the deliberate disregard of in principle available information, as they are in classical statistical physics. Even if all relevant facts are known---even if there is nothing relevant to be ignorant of---the outcome of a measurement cannot in general be predicted with certainty. The residual probabilities, which remain when all relevant facts are taken into account, may thus be considered objective. (Once we have obtained more clarity regarding our epistemological situation, I will check back on how this view fares.)

To QBists, on the other hand, the measurement outcomes to which quantum mechanics serves to assign probabilities are \textit{as subjective as} the probabilities that are assigned to them; they are private, subjective experiences. There seems to be only one way to make sense of this startling view, to wit: as a reaction to the charge of instrumentalism. To many physicists familiar with it, QBism does not significantly differ from instrumentalism, the view that quantum mechanics is concerned with nothing but statistical correlations between the readings of measuring instruments. Treating the readings of measuring instruments as private experiences wards off the charge of instrumentalism but invites the charge of solipsism instead.

A minimal instrumentalism is part and parcel of any physical interpretation of the mathematical formalism of the quantum theory. What remains open to debate is the epistemological or ontological status of the events to which the theory serves to assign probabilities. The claim that they are subjective experiences does not ring true, but neither does full-fledged instrumentalism, according to which the measurement apparatus is part of a world that (a)~exists independently of perceiving and thinking subjects and (b)~has the very properties it has in our perception-informed thoughts. Because we do not have the slightest idea of what a measurement apparatus (or any other object for that matter) is like in itself, the notion that it is in itself (more or less) as it is in our perception-informed thoughts is, to say the least, unwarranted.

The fact that we all experience the same objective world (albeit from different vantage points) is generally ``explained'' by postulating a world of real objects which are the causes of sense-impressions, and which produce roughly the same impressions on all of us. But this explains nothing, for while we can compare our respective experiences by communicating with each other, we have no way of comparing our sense-impressions with the objects that are invoked as their causes.

What QBists are actually trying to defend is a \textit{more sensible} conception of the objective world, which takes account of the fact that it exists \textit{for us}, and that it is what it is not in itself but because of the way it is perceived and conceived by us. Such a conception was first put forward by Immanuel Kant and later, in the context of quantum mechanics, by Niels Bohr. To Kant as well as to Bohr, the objective world is something we construct on the basis of our private experiences, using concepts that owe their meanings to (a)~the logical structure of human thought (or else the grammatical structure of human language) and (b)~the spatiotemporal structure of human sensory experience. It is because we have these structures in common that we are able to understand each other whenever we use those concepts.

Mermin invokes the celebrated probabilist Bruno de Finetti, who wrote: ``The abandonment of superstitious beliefs about the existence of Phlogiston, the cosmic ether, absolute space and time\dots, or Fairies and Witches, was an essential step along the road to scientific thinking. Probability too, if regarded as something endowed with some kind of objective existence, is no less a misleading misconception, an illusory attempt to exteriorize or materialize our actual probabilistic beliefs.'' 

Taking the mind-independent existence of the external world for granted, de Finetti holds that there is no place for probability \textit{in such a world}, anymore than there is for Phlogiston and the rest. This may be true, but it is beside the point. It is an essential step \textit{in the wrong direction} to assume that science is concerned with a mind-independently existing world. What Kant has taught us, what Schr\"odinger\cite{SchroedingerPO} has stressed, and what was always at the back of Bohr's mind,\cite{Mohrhoff2020} is that the empirically accessible world is a world that we collectively construct using concepts that owe their meanings to the logical structure of human thought and the spatiotemporal structure of human sensory experience.

I wish QBists were always as clear as Mermin when he stresses the distinction between a weak and a strong objectivity (to use the terminology of d'Espagnat\cite{dEspagnat68}): ``Those who reject QBism \dots\ reify the common external world we have all negotiated with each other, purging from the story any reference to the origins of our common world in the private experiences we try to share with each other through language''.\cite{MerminQBnotCop} The weakly objective world is the one we have all negotiated with each other; a strongly objective world (if it were a consistent notion) would be the same world reified, purged of its origins in our private experiences.

\looseness=1 The most important among the aforementioned concepts are ``substance'' and ``causality.'' Of the former there are essentially two varieties: Lockean and Aristotelian. A Lockean substance serves as the “glue” that binds properties in the manner in which a single logical (or grammatical) subject bundles predicates. It represents the \textit{thought} that several properties are properties of one and the same object. An Aristotelean substance betokens the self-existence of a bundle of properties. Here the idea is that properties enjoy actual existence only if they are possessed by, or attributable to, a substance. There is an important caveat, though: by attributing independent existence to an object, we do not imply that the object exists independently of ourselves qua thinking and perceiving subjects. What is meant is that we can talk or more generally behave \textit{as if} the object enjoyed a mind-independent existence. What distinguishes quantum physics from everything that preceded it, is that it does not allow us to be that cavalier about our part in the creation of our common external world. The weak objectivity of the quantum world (established by Bohr) is considerably weaker than the weak objectivity of the classical world (established by Kant).

To see this clearly, it helps to consider the difference between an individual quantum object and the type or species to which it belongs. In her monograph \textit{Particle Metaphysics},\cite{Falkenburg2007} Brigitte Falkenburg has characterized quantum objects (qua types) as “Lockean empirical substances, that is, collections of empirical properties which constantly go together.” These “are only individuated by the experimental apparatus in which they are measured or the concrete quantum phenomenon to which they belong.” They ``propagate through an experimental context, respecting the conservation laws of mass–energy, charge, and spin (as well as Einstein's causal condition that signal transmission is only possible within the light cone)'' not on the backs of property-carrying substances, but as collections of empirical properties that ``exhibit non-local, acausal correlations'' (between events in a classical world). Subatomic reality, therefore, is a top-down construct: “The opposite bottom-up explanation of the classical macroscopic world in terms of electrons, light quanta, quarks, and some other particles remains an empty promise.”

Whereas classically one could imagine a property-carrying substance bestowing actual and (for all practical purposes) mind-independent existence on the properties it carries, in the case of a quantum object the carrier of properties ``is nothing but the quantum phenomenon itself, say, a particle track on a bubble chamber photograph or the interference pattern of the double slit experiment.'' But if the properties of a quantum object owe their actual existence to an apparatus or a quantum phenomenon, then the properties of the apparatus or quantum phenomenon cannot owe \textit{their} existence to the atoms or subatomic particles of which they are commonly said to be composed. What, then, accounts for their existence? According to QBists, their being experienced. But this is only half of the story. The other half accounts for their \textit{objective} existence. And this half we owe to Bohr\cite{Bohr61}:
\begin{quote}
From our present standpoint physics is to be regarded not so much as the study of something a priori given, but rather as the development of methods for ordering and surveying human experience. In this respect our task must be to account for such experience in a manner independent of individual subjective judgement and therefore objective in that sense, that it can be unambiguously communicated in the common human language. 
\end{quote}
What accounts for the objective existence of the properties of classical objects is (a)~that they are experienced and (b)~that we can \textit{talk} about them \textit{as if} they existed independently of human conscious experience. And that would be the end of the story if there were no quantum objects, or if it were not necessary to distinguish between a classical domain containing classical objects and a quantum domain containing quantum objects, or if the properties of quantum objects were not contextual, i.e., defined in terms of classical objects and dependent for their existence on the experimental conditions in which they are observed. For the click of a counter does not simply indicate the presence of something inside the region monitored by the counter. Instead, the counter \textit{defines} a region, and the click \textit{constitutes} the presence of something inside it. Without the click, \textit{nothing} is there, and without the counter, there is no \textit{there}.

What essentially distinguishes Bohr's thinking from Kant's is his insight that the spatiotemporal resolution of sensory experience is limited by Planck’s constant. If sensory experience could in principle reach ``all the way down,'' we could detach ourselves from our common external world in the manner first conceived by Kant. But since we need to distinguish between two categories of objects, and since the distinction involves the reach of human sensory experience---classical objects being directly accessible to it, quantum objects being beyond the reach of direct sensory experience---this possibility no longer exists.

If the spatiotemporal resolution of sensory experience is limited, and if the construction of the objective world depends on concepts that owe their meanings in part to the spatiotemporal structure of sensory experience, then the concepts needed to construct the objective world can only be employed---and the construction of the objective world can only be carried---as far as sensory experience can reach. The only meaningful way of talking about quantum objects, therefore, is to talk (a)~about the experimental arrangements by which they are prepared, created, detected, or observed, and (b)~about the statistical correlations that obtain between preparations/creations and detections/observations. All of this was clearly seen by Bohr\cite{Bohring}:
\begin{quote}
[T]he facts which are revealed to us by the quantum theory ... lie outside the domain of our ordinary forms of perception [BCW\,6:\,217].

\medskip [I]t is decisive to recognize that, \textit{however far the phenomena transcend the scope of classical physical explanation, the account of all evidence must be expressed in classical terms}. The argument is simply that by the word ``experiment'' we refer to a situation where we can tell others what we have done and what we have learned and that, therefore, the account of the experimental arrangement and of the results of the observations must be expressed in unambiguous language with suitable application of the terminology of classical physics.\cite{Bohr58}

\medskip [T]he quantum-mechanical formalism \dots\ represents a purely symbolic scheme permitting only predictions \dots\ as to results obtainable under conditions specified by means of classical concepts. [BCW\,7:\,350--351]
\end{quote}
By ``classical concepts'' Bohr did not mean concepts proprietary to classical physics but concepts which owe their meanings to our ``forms of perception,'' and by this he meant ``the conceptual structure upon which our customary ordering of our sense-impressions depends and our customary use of language is based''[BCW\,10:\,xxxv--xxxv].

As to the second of the two cornerstones of classical discourse---causality---it derives its meaning (a)~from the logical relation between an antecedent (``if\dots'') and a consequent (``then\dots'') and (b)~from the temporal relation between ``now'' and ``then'': if this happens or is the case \textit{now} then that will happen or be the case \textit{then}. The reason this concept has no meaningful application to the quantum domain is not merely the statistical nature of quantum-mechanical predictions but more importantly the twofold conditionality of the predictions: if$_1$ such and such quantum state has been prepared, and if$_2$ such and such measurement will be made at such and such time, then these are the probabilities of the possible outcomes\dots.

On the other hand, if the concepts of substance and causality had no meaningful application to the classical domain---in which case there would be no classical domain---we would not be in a position to describe the experimental arrangements by which quantum objects are prepared, created, detected, or observed. Quantum mechanics itself would have no domain of application, for it would be impossible to talk about the events to which quantum states serve to assign probabilities. This is why Bohr,\cite{vW2006} responding to the suggestion that classical concepts may one day be replaced by quantum theoretical ones, said: ``We also might as well say that we are not sitting here and drinking tea but that all this is merely a dream.''

\section{Conclusion}
Quantum theory has revealed a fundamental limit to the spatiotemporal resolution of sensory experience. The first consequence of the existence of this limit is that the concepts needed to construct the objective world---in particular the concepts of ``substance'' and ``causality''---can only be employed as far as sensory experience can reach. Several corollaries follow, foremost among them the need to distinguish between two categories of objects: classical objects, which are directly accessible to sensory experience, and quantum objects, which are beyond the reach of direct sensory experience. As a result, the properties and behaviors of quantum objects are contextual, i.e., they are defined in terms of the properties and behaviors of classical objects, and they depend for their existence on the experimental conditions in which they are observed. Hence the only meaningful way of speaking about quantum objects is to speak about (a)~the experimental arrangements by which they are prepared, created, detected, or observed and (b)~the statistical correlations that obtain between preparations/creations and detections/observations. This is why ``the physical content of quantum mechanics is exhausted by its power to formulate statistical laws governing observations obtained under conditions specified in plain language'' [BCW\,10:\,159].

The weak objectivity of Kant's world made it possible to talk and otherwise behave \textit{as if} objects of sensory experience existed independently of sensory experience. The existence of a fundamental limit to the spatiotemporal resolution of sensory experience implies a further weakening of the objectivity of the objective world. It requires us to make a distinction between two categories of objects---a distinction that is predicated on a fundamental limitation of (human) \textit{sensory experience}. Even those who wish to abolish the fundamental difference between quantum objects and measuring instruments are forced to acknowledge that \textit{we} need the latter for the purpose of \textit{observing} the former, or that the purpose of the latter is to \textit{indicate} to \textit{us} properties of the former.

As long as the reach of sensory experience was in principle unlimited, the elision of the knowing subject from the objects known was as painful (or not) as separating fraternal twins. When quantum mechanics came along, it became as painful (even to the onlooker) as separating conjoined twins. The least painful way to suppress our involvement in the creation of an objective world is a metaphysically sterile instrumentalism, which reduces talk about atoms and subatomic particles to correlations between the behaviors of measuring instruments. Any way of talking about quantum objects \textit{per se} is at once confronted with the contextuality of their properties and with the resulting circularity of accounting for the properties of classical objects in terms of the properties of quantum objects. 

Tellingly, $\Psi$-ontology, which reifies wave functions, ends up performing the same vanishing trick with regard to quantum objects as does (full-fledged) instrumentalism, which reifies classical objects.%
\footnote{When physicists reflect on the motives for their research, they nonetheless often claim (especially on TV, in press releases, and in grant applications) that their aim is to discover the elementary building blocks of the universe and the processes by which these interact with each other. Go figure.}
Thus Sheldon Goldstein writes:\cite{GoldsteinSEP} 
\begin{quote}
It seems clear that quantum mechanics is fundamentally about atoms and electrons, quarks and strings, not those particular macroscopic regularities associated with what we call measurements of the properties of these things. But if these entities are not somehow identified with the wave function itself---and if talk of them is not merely shorthand for elaborate statements about measurements---then where are they to be found in the quantum description?
\end{quote}
Since neither instrumentalists nor $\Psi$-ontologists have anything illuminating to say about atoms and fundamental particles, it may fall to QBists to do so, and maybe even to reinvigorate the objectivity of our common external world in doing so.\cite{Mohrhoff2022}

I promised to check back on the notion that the residual probabilities, which remain when all relevant facts are taken into account, may be regarded as objective. In view of the utter weakness of what is presently defendable as objective reality, that notion has nothing whatsoever to commend it.

One final point. Since the time of Bishop Berkeley, philosophers have been asking themselves: if a tree falls in a forest and no one is around to hear it, does it make a sound? By the same token one may ask: if a measurement is made and no one is around to observe the outcome, does it have an outcome? The instrumentalist will insist that of course it does. The QBist will insist that of course it does not. The $\Psi$-ontologist will hem and haw eloquently. What might Bohr have said? 

\looseness=1 If Alice and Bob (or Wigner and his friend\cite{Wigner61}) are honest reporters of their respective experiences, and if Alice tells Bob that she has obtained an outcome without telling him which, then Bob will have to treat Alice's possible outcomes as elements of a mixture (of quantum states) rather than as elements of a superposition. But that is the least of it. A measurement apparatus is something that, among other things, (a)~is accessible to direct sensory experience, and (b)~is capable of indicating not only \textit{which} possible outcome has been obtained but also \textit{that} an outcome has been obtained. If, then, we can agree that measurement apparatuses exist---we are no obligation to explain this obvious fact, anymore than we are obliged to explain why there is anything rather than nothing at all---we can also agree that an outcome exists whenever a measurement apparatus indicates that that is the case, whether or not anyone is around to take cognizance of the outcome. The reason is that the world in which measurements take place has been constructed by us---its existence is predicated on us qua perceiving and thinking subjects, rather than on the intrinsic reality of ultimate constituents---and so there is no need to invoke observers a second time.

\looseness=1 This should also put paid---though of course it won't---to the notion that quantum states were collapsing in the early universe, long before there were any physicists. Before we arrived on the scene, the common external world we have all negotiated with each other did not exist. The objective world (including its past) is something that humans have constructed on the basis of experiences accumulated over a rather brief period of history. I am reminded here of a point that has been made by Yale historian Timothy \hbox{Snyder}, most recently in his lecture series ``The Making of Modern Ukraine''.\cite{Snyder} It is that the genesis of a nation as told by that nation's historians tends to differ markedly from the genesis of the same nation as told by other historians. The former usually has a legendary flavor, as contrasted with the more objective flavor of the latter. Unfortunately, when it comes to telling the genesis of our universe, only the former option is available.

\end{document}